\documentclass[twocolumn]{jetpl}
\input epsf
\usepackage{epsfig}

\def\ltap{\raisebox{-.55ex}{\rlap{$\sim$}} \raisebox{.4ex}{$<$}}

\def\lsim{\mathrel{\ltap}}

\newcommand{\be}{\begin{equation}}
\newcommand{\ee}{\end{equation}}
\newcommand{\ba}{\begin{array}}
\newcommand{\ea}{\end{array}}
\def\simleq{\; \raise0.3ex\hbox{$<$\kern-0.75em \raise-1.1ex\hbox{$\sim$}}\; }
\def\simgeq{\; \raise0.3ex\hbox{$>$\kern-0.75em \raise-1.1ex\hbox{$\sim$}}\; }
\newcommand{\eV}{{\rm eV}}
\newcommand{\Mpc}{{\rm Mpc}}
\newcommand{\kpc}{{\rm kpc}}
\newcommand{\muG}{\mu{\rm G}}

\lat

\title{Mapping deflections of extragalactic Ultra-High Energy Cosmic
     Rays in magnetohydrodynamic simulations of the Local Universe}
\rtitle{Mapping deflections of extragalactic UHECR in simulated Local Universe}
\sodtitle{Mapping deflections of extragalactic Ultra-High Energy Cosmic
     Rays in magnetohydrodynamic simulations of the Local Universe}

\author{ Klaus Dolag$^a$, Dario~Grasso$^b$, Volker~Springel$^c$ and
Igor Tkachev$^d$}
\address{$^a${\it Dipartimento di Astronomia, Universit\`a di Padova,
Padua, Italy\\}
$^b${\it Scuola Normale Superiore and I.N.F.N., Pisa, Italy\\}
$^c${\it Max-Planck-Institut f\"ur Astrophysik, Garching, Germany\\}
$^d${\it CERN - Theory Division, Geneve, Switzerland}}

\rauthor{K. Dolag, D. Grasso, V. Springel, I. Tkachev}
\sodauthor{K. Dolag, D. Grasso, V. Springel, I. Tkachev}

\abstract{
We construct a map of deflections of ultra-high energy cosmic rays by
extragalactic magnetic fields using a magneto-hydrodynamical
simulation of cosmic structure formation that realistically reproduces
the positions of known galaxy clusters in the Local Universe. Large
deflection angles occur in cluster regions, which however cover only
an insignificant fraction of the sky. More typical deflections of
order $\lsim 1^\circ$ are caused by crossings of filaments. For
protons with energies $E \geq 4 \times 10^{19}\,{\rm eV}$, deflections
do not exceed a few degrees over most of the sky up to a propagation
distance of 500 Mpc. Given that the field strength of our simulated
intergalactic magnetic field forms a plausible upper limit, we
conclude that charged particle astronomy is in principle possible.
}


\begin{document}

\maketitle

\paragraph*{Introduction.}

Considerable effort is presently undertaken around the world to build
\cite{Auger,EUSO} experiments devoted to determining the composition, the
energy spectrum and the arrival directions of Ultra High Energy Cosmic Rays
(UHECR). This challenge is in part motivated by the Greisen-Zatsepin-Kuzmin
(GZK) puzzle~\cite{GZK} which became particularly acute with Fly-Eye and AGASA
data \cite{experiments}, and by the realization that the UHECR flux at $E >
10^{19}~\eV$ is probably dominated by the emission of sources which are quite
different from conventional galactic sources.  The directional information may
allow the identification of UHECR sources, provided primary particles are not
deflected too much by galactic and intergalactic magnetic fields
(IGMFs). 

Several arguments suggest that UHECR are electrically charged nuclei, most
probably they are protons. The possibility of neutral particles is not ruled
out, but needs not be discussed here since such rays point back to the sources
anyway. It is possible that a fraction of UHECR is comprised of iron nuclei,
see e.g.~Ref.~\cite{Mikhailov:2003at}. However, according to an analysis of
inclined events recorded by the Haverah Park shower detector
\cite{Ave:2000nd}, above $10^{19}$ eV less than 30\% of the primary cosmic
rays can be iron nuclei at the 95\% confidence level. In what follows we
normalize our results to the case of protons. The case of other nuclei can be
recovered by multiplication with their charge, Z.

Galactic magnetic fields (MF) with $B_{\rm gal} \sim 1~\muG$ are not expected
to produce significant deflections at extremely high energies, $E \simgeq
10^{20}~\eV$ in the case of protons. Even at lower energies $E \sim 4\times
10^{19}~\eV$, strategies have been proposed which allow source identification
without detailed knowledge of the galactic MF~\cite{Tinya:03}.

The very attractive perspective to do astronomy with proton primaries might
however be spoiled by the presence of strong IGMFs. So far, evidences of the
presence of IGMFs have been found only within, or very close to, rich clusters
of galaxies. The most relevant observations are those based on Faraday
rotation measurements (RM) of the polarized radio emission of sources located
within or behind clusters, and on the synchrotron emission of relativistic
electrons in the intracluster MF. The results of both methods imply the
presence of MFs with strength at the $\muG$ level extending up to 1 Mpc from
cluster centers. The coherence length of the field is inferred to lie in the
range $10-100~\kpc$ (see recent review~\cite{Carilli} and references
therein). Such fields do certainly induce large deflections of UHECR protons
that cross clusters of galaxies. However, galaxy clusters fill only a tiny
fraction of the volume of the universe, so that we may expect them to produce
large deflections at best over a small portion of the sky
\cite{Berezinsky,Blasi:03}.  Outside clusters, only upper limits on the IGMF
strength are available. They are at the level of $1-10$ nG for fields
extending over cosmological distances with coherence lengths in the range 50
to 1 Mpc, respectively~\cite{Blasi:99}. These bounds do not hold for MFs in
clustered regions, like filaments connecting galaxy clusters where the field
might be as large as $10^{-7}~{\rm G}$.  In principle, either a weak all
pervading smooth field, or stronger fields localized in a complex web of
filaments, may produce sizable deflections of UHECR over a large portion of
the sky.  It is hence evident that a better knowledge of the large-scale
magnetic structure of the universe is called for.

In this letter, we approach this problem by means of numerical
simulations of cosmic structure formation, where we combine the
collisionless dynamics of the dark matter component with the
magnetohydrodynamics (MHD) of the magnetized gas. Our basic hypothesis
is that the MFs observed in rich clusters of galaxies are the outcome
of a MHD amplification process powered by the hierarchical formation
of clusters. This assumption is supported by the results of previous
simulations which, under the same hypothesis that we adopt here,
succeeded to reproduce the general features of RM in several observed
clusters \cite{Dolag:2002}. The tiny seed field required to initiate
the amplification process may be either of primordial
origin~\cite{report} or the result of a battery associated with the
initial stages of structure formation \cite{Kulsrud:1996km}.

Simulations of the magnetic structure of the universe and of the UHECR
propagation within it have been previously attempted by several
authors \cite{Kulsrud:1996km,Ryu:1998,Sigl:2003a}. A novel achievement
of our work is that we have performed {\it constrained simulations},
which reproduce the observed large scale structures in the nearby
universe, leaving essentially no ambiguity for the choice of observer
position.  This is quite relevant in the present context, since it has
been shown [\cite{Sigl:2003a}] that the angular distribution, as well
as the energy spectrum, of UHECRs reaching an observer located in a
weakly magnetized region may differ considerably from that seen by a
strongly magnetized one.  Furthermore, by tracing UHECR trajectories
in the simulated magnetic structures we are able to construct maps of
expected UHECR deflections as a function of distance that, for the
first time, account for the actual large-scale structure as seen from
the Galaxy.

\paragraph*{MHD simulations of the Local Universe.}

We use initial conditions that were constructed from the IRAS 1.2-Jy
galaxy survey by first smoothing the observed galaxy density field on
a scale of 7 Mpc, evolving it linearly back in time, and then using it
as a Gaussian constraint for an otherwise random realization of the
$\Lambda$CDM cosmology. In \cite{Mathis} it was shown that these
constrained initial conditions, when evolved to the present time,
reproduce the observed density and velocity field of the Local
Universe. In addition, they allow a direct identification of prominent
clusters (Virgo, Coma, Centaurus, Hydra, Perseus, A3627, and Pavo)
with counterparts formed in the simulation, which are found at the
right places, and with approximately the correct observed masses.  We
extended the initial conditions of~\cite{Mathis} by adding gas,
together with an initial MF. The volume filled by high resolution
particles within our simulation is a sphere of radius $\sim 115$ Mpc
centered on the Milky Way.  This region comfortably includes the
entire Local Super Cluster (LSC) and is modelled with a maximum
spatial resolution of $10\,{\rm kpc}$. The simulation uses 51 million
gas particles of mass $6.9\times 10^8\,{\rm M}_\odot$, 51 million
high-resolution dark matter particles, and an additional 7 million
boundary particles in the distant low-resolution region.

We evolved the initial conditions with the newest version of the
{\small GADGET}-code \cite{springel}, adding the Magnetic Smoothed
Particle Hydrodynamics (MSPH) technique~\cite{Dolag:2002} to follow MF
evolution.  Previous work \cite{Dolag:2002} showed that magnetic seed
fields in the range of $(1-5) \times 10^{-9}\,{\rm G}$ at redshift
$z_* \simeq 20$ will be amplified due to the structure formation
process and reproduce RM in clusters of galaxies. This corresponds to
$B(z_{*}) (1 + z_{*})^{-2} \simeq 0.2-1 \times 10^{-11}$ G at the present
time in the unclustered intergalactic medium (IGM). It was also
demonstrated that the MF amplification process completely erases any
memory of the initial field configuration in high density regions like
galaxy clusters.  Therefore, we can safely set the coherence length
$l_c(z_{\rm in})$ of the initial seed field to be infinite in our
simulation. Although this assumption is probably unrealistic, it does
not lead to underestimation of the UHECR deflections.  Concerning the
initial strength of the MF, we used the highest value which still
allowed previous MSPH simulations to successfully reproduce RM in
clusters, i.e.~the results presented here give safe upper bounds on
UHECR deflections.

\begin{figure*}
\includegraphics[width=0.95\textwidth]{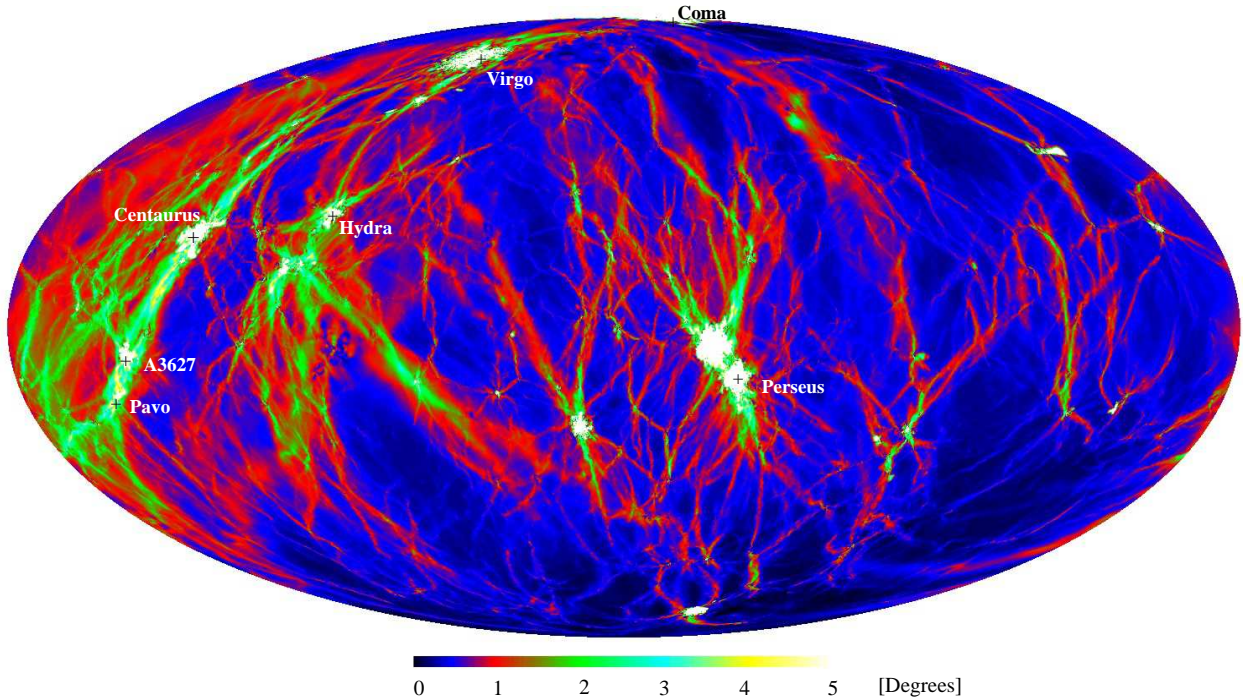}
\caption{FIG. 1. ~Full sky map (area preserving projection) of deflection
angles for UHECRs with energy $4\times10^{19}$ eV using a linear color
scale. All structure within a radius of $107\,$ Mpc around the
position of the Galaxy was used. The coordinate system is galactic,
with the galactic anti-center in the middle of the map.  Positions of
identified clusters are marked using the locations of the
corresponding halos in the simulation. Note that deflections internal
to the Milky Way have not been included.}
\label{map}
\end{figure*}

Clusters are generally connected by magnetized filamentary structures
of gas and dark matter, where high-density filaments often harbor
small clusters or groups. We find that shock fronts and shear flows
are ubiquitous in these filaments, giving rise to substantial MHD
amplification in these structures, boosting the MF intensity far above
the expectation of adiabatic compression alone, as pointed out in
previous work \cite{Dolag:2002}.  We also identified low density
filaments where the MF is roughly aligned along their axis, with a
strength of $\sim 10^{-4}~ \muG$. This is consistent with a purely
adiabatic amplification of the seed MF due to the compression of field
lines.  We find no significant MF in the neighborhood of the Milky
Way's position, which is contrained to lie within a sphere of 7 Mpc
around the origin. We find a group of four halos aligned within the
super galactic plane in this region, corresponding to the Local
Group. Due to the lack of small-scale constraints it is not certain
which of the four galaxies should be best associated with the Milky
Way, but this does not affect our results. Because it is a small
galaxy group, the MFs associated with the group are sufficiently
weak to not lead to significant deflections, despite covering a large
fraction of the sky.

\paragraph*{Deflections of charged UHECR.}

Having obtained a 3D model of MFs in the Local Universe, we can
construct an associated map of deflections of charged particles under
the action of the Lorentz force. We here consider only protons with
energy $E = 4 \times 10^{19}~ \eV$. This is the threshold value for
the process of photo-pion production in collisions with Cosmic
Microwave Background (CMB) photons ($p + \gamma_{CMB} \rightarrow p(n)
+ \pi^{0(+)}$).  The energy loss length is large, $l_E \sim 1000~\Mpc$
(for a recent review see e.g.~\cite{Ancho:02}), and initially higher
proton energies quickly degrade into this range. Neglecting energy
losses and taking $E = 4 \times 10^{19}~ \eV$ to be the energy at
detection, we obtain upper bounds for the deflections of protons with
higher energy since the deflection angle decreases linearly increasing
the energy.

We do not follow particle trajectories directly; instead we compute
accumulated deflections along rectilinear paths. This is a reasonable
simplification since we are not interested in actual source positions,
but rather in finding directions with small deflections.  In
Fig.~\ref{map}, we show a deflections map obtained by tracing an
isotropic distribution of protons from a maximal distance of $d_{\rm
max} = 107~\Mpc$ to the observer.  
Recall that Fig.~\ref{map} represents a map of
deflections, not a distribution of arrival directions. The former
is independent of the assumed distribution of UHECR sources.

The pattern of clusters and filaments is clearly visible in
Fig.~\ref{map}.  Large deflections are produced only when protons
cross the central regions of galaxy clusters, and most of these
strongest deflections are found along a strip which can be
approximately identified with the Great Attractor. The observed
positions of Virgo, Coma, Hydra and Centaurus lie in this
region. Their locations quite precisely coincide with regions where
the deflections exceed $4^\circ$. Perseus and other minor clusters
produce large deflections in other well delineated regions of the
sky. Outside clusters, which occupy only a small fraction of the sky,
deflections of $1^\circ- 2^\circ$ occur along an intricate network of
filaments, covering a larger area. The regions with $\delta \ll
1^\circ$ correspond to voids where the MF strength is even smaller
than $10^{-11}$ G.

In order to investigate the relative importance of deflectors at different
distance, we also produced deflection maps that only included
deflectors up to some maximum distance.  We observe no significant
deflections produced at distances smaller than 7 Mpc. 
Massive clusters at large distances ($\sim 100~$ Mpc) produce
large deflections but cover only a negligible fraction of the sky, so
that the bulk of the deflections is produced by passages through
filaments.

\begin{figure}
\includegraphics[width=0.475\textwidth]{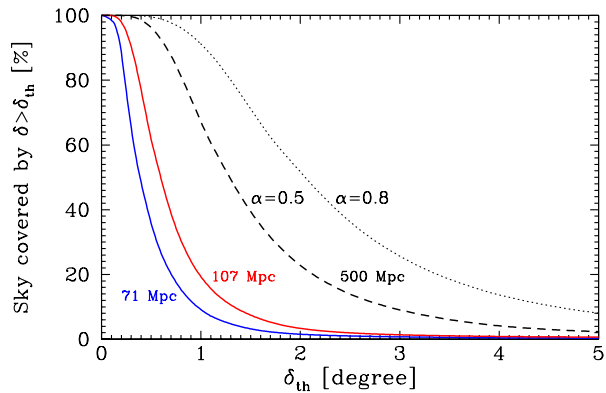}
\caption{FIG. 2. ~Cumulative fraction of the sky with deflection angle larger
than $\delta_{\rm th}$, for several values of propagation distance
(solid lines). We also include an extrapolation to $500\,$ Mpc, assuming self 
similarity with $\alpha=0.5$
(dashed line) or $\alpha=0.8$ (dotted line). The assumed UHECR energy
for all lines is $4.0\times10^{19}$ eV.}
\label{histo}
\end{figure}

In Fig.~\ref{histo}, we plot the fraction of the sky, $A(\delta_{\rm
th})$, over which deflections larger than $\delta_{\rm th}$ are found,
for different propagation distances. We see that deflections larger
than $1^\circ$ are to be expected over less than $20\%$ of the sky up
to the distance $d = 107~\Mpc$. For large distances
$d$, we find that $A(\delta_{\rm th},d)$ approaches a self-similar
behavior, viz.~$A(\delta_{\rm th},d) = A_0 (\delta_{\rm th}\times
(d_0/d)^\alpha)$.  Numerically, we observe $\alpha = 0.8$ for $70 <
d/\Mpc < 110$. Self-similarity is consistent with the assumption that
the density of deflectors (filaments) reaches a constant value at
large distances. Since MFs are uncorrelated in different filaments,
multiple filament crossings should produce a ``random walk'' in the
deflection angle, resulting in $\alpha = 0.5$. 
The value of $\alpha = 0.8$ we observe may hence indicate that the regime of
multiple filament crossings is not yet reached over the distances
probed by our simulation. We include an extrapolation of
$A(\delta_{\rm th},d)$ up to a distance of $500~\Mpc$
in Fig.~\ref{histo}, shown for two values of $\alpha$, the observed
one of $\alpha =0.8$, and the expected one for large propagation
distances, $\alpha =0.5$. We expect that these two curves bracket the
range of true deflections at $E \sim 4 \times 10^{19}~\eV$.
 
We comment finally on the potential effect of the
unclustered component of the IGMF, i.e.~the field in voids and low
density regions outside of clusters and filaments, coherence length
of which, $l_c$, is unconstrained by our simulations. 
If $l_c \ll d$ the proton trajectory makes a random walk through the
magnetic domains, and the overall deflection is given by
$$
 \delta \simeq 0.2^o\;
\left(\frac{B_0}{E}\frac{4\times 10^{19}~\eV}{10^{-11}~{\rm
G}}\right) \left(\frac{d}{1~{\rm Gpc}}\right)^\frac {1}{2}
\nonumber\left(\frac{l_c}{1~\Mpc}\right)^\frac {1}{2}.
$$ Hence, observable deflections are not produced by the unclustered
component of the IGM if $l_c$ is smaller than a few tens of Mpc. Note
that such small coherence lengths are expected from most of the
proposed generation mechanisms of seed IGMFs \cite{report}. 
The few mechanisms predicting larger $l_c$ generally give rise to MF which 
are too weak to produce observable deflections of UHECR.  
Furthermore, an unclustered IGMF does not exist at all
if the seed field is generated by a battery powered by structure
formation \cite{Kulsrud:1996km}.

\paragraph*{Conclusions.}

We presented the first map of UHECR deflections in the Local Universe
that is based on a simulation that realistically reproduces the known
large-scale structure around the Galaxy, while simultaneously
following the MHD amplification of MFs during cosmic structure
formation. The positions and masses of the most prominent clusters are
reproduced well in our simulation.  This is an important advantage of
our technique. Since local structures subtend large angles on the sky,
it is important to be able to reliably identify ``bad'' regions of
expected large deflections, a task that can be accomplished using our
map, thereby providing important guidance for UHECR source
identification.  Provided our basic hypothesis about the origin of
IGMF is correct, our results should be understood as upper bounds for
the expected deflection angles, because we have used the largest seed
field still compatible with the RM in clusters, and secondly, we
neglected UHECR energy losses on the path to the detector.  The actual
observation of stronger deflections would imply that the evolution of
the IGMF is not merely passive, possibly indicating a pollution of the
IGM by physical process such as galactic winds.

We have also extrapolated the distribution of deflection angles to very large
source distances in a statistical manner. Out to 500 Mpc and at $E \geq 4
\times 10^{19}~{\rm eV}$, typical deflections are smaller than the angular
resolution of current ground array UHECR detectors over more than half of the
sky (but may exceed angular resolution of stereo fluorescent detectors).  This
result is consistent with an observed small-scale clustering of UHECR arrival
directions \cite{CRclusters}
\footnote{There is no clustering in the current HiRes data
\cite{Abbasi:2004ib,Abbasi:2004dx}, which became avaliable after submission of
our paper. However, with the current statistics there is no contradiction yet
\cite{Abbasi:2004dx,Yoshiguchi:2004np}.}  and with evidence for a BL Lacs -
UHECR correlation \cite{Tinyakov:2001nr} in the energy range $E \sim 4 \times
10^{19}~{\rm eV}$ being due to protons \cite{Tinyakov:2001ir}.  On the other
hand, our results do not support models which invoke strong MFs in the local
universe to solve the GZK anomaly as well as models which explain small-scale
clustering by magnetic lensing.

We conclude that charged particle astronomy should in principle be
possible regardless of the way the GZK problem will be resolved.

\section*{Acknowledgments}
{\tolerance=400 We thanks V. Berezinsky, P. Blasi, D. Semikoz, 
P. Tinyakov, M. Vietri for reading the manuscript and providing useful 
comments. The simulations were carried out on the IBM-SP4
machine at the `Rechenzentrum der Max-Planck-Gesellschaft', with
CPU time assigned to the `Max-Planck-Institut f\"ur Astrophysik'.
Fig. \ref{map} has been produced using HEALPix \cite{healpix}.
K.~Dolag acknowledges support by a Marie Curie Fellowship of the
European Community program `Human Potential' under contract number
MCFI-2001-01227.}

\end{document}